# Molecular Thin Films: a New Type of Magnetic Switch

Dr Sandrine Heutz[1, 2, *], Dr Chiranjib Mitra[2], Dr Wei Wu[2], Prof Andrew J. Fisher[2], Dr Andrew Kerridge[2, ◊], Prof Marshall Stoneham[2], Dr Tony H. Harker[2], Miss Julie Gardener[2], Miss Hsiang-Han Tseng[3], Prof Tim S. Jones[3,†], Prof Christoph Renner[2,‡], and Prof Gabriel Aeppli[2]

[1] Department of Materials and London Centre for Nanotechnology,
Imperial College London, London SW7 2AZ, UK.
[2] Department of Physics and Astronomy and London Centre for Nanotechnology,
University College London, London WC1E 6BT, UK.
[3] Department of Chemistry and London Centre for Nanotechnology
Imperial College London, London SW7 2AZ, UK.
[◊] Present address: Department of Chemistry, University College London,
London WC1E 6BT, UK.
[†] Present address: Department of Chemistry, University of Warwick,
Coventry CV4 7AL, UK.
[‡] Present address: DPMC–MaNEP, University of Geneva, 24 Quai Ernest-Ansermet,
CH-1211 Geneva 4, Switzerland.
[*] E-mail:s.heutz@imperial.ac.uk; fax: +44 (0) 207 594 6757

**The design and fabrication of materials that exhibit both semiconducting and magnetic properties for spintronics [1] and quantum computing has proven difficult. Important starting points are high-purity thin films as well as fundamental theoretical understanding of the magnetism. Here we show that small molecules have great potential in this area, due to ease of insertion of localised spins in organic frameworks and both chemical and structural purity. In particular, we demonstrate that archetypal molecular semiconductors, namely the metal phthalocyanines (Pc), can be readily fabricated as thin film quantum antiferromagnets, important precursors to a solid state quantum computer. Their magnetic state can be switched via fabrication steps which modify the film structure, offering practical routes into information processing. Theoretical calculations show that a new mechanism, which is the molecular analogue of the interactions between magnetic ions in metals, is responsible for the magnetic states. Our combination of theory and experiments opens the field of organic thin film magnetic engineering.**

*Advanced Materials (re-submitted 17-08-07)*Phthalocyanines (abbreviated Pc for the phthalocyanato ion $C_{32}H_{16}N_8^{2-}$) are polyaromatic molecules which can accommodate a range of atoms or groups of atoms (in the 2+ oxidation state) in their central cavity, as shown in Figure 1.a.[2] They have now become archetypal organic semiconductors, and their attractive optoelectronic properties are already being extensively exploited in solar cells,[3] field effect transistors and light emitting diodes.[4] Magnetic studies have been comparatively sparse, and have mainly focused on MnPc single crystals, which exhibit ferromagnetism below about 10K.[5] The influence of crystal structure on magnetic properties has recently become apparent via pressure-dependent investigations,[6] dopant-induced crystal modification,[7] and film studies.[8]

One of the important long-term demands on spintronics is for quantum computing, which needs useful spin entanglement. Entanglement in ordered solids is most easily achieved for low dimensionality, low spin ions and antiferromagnetic coupling. These conditions are fulfilled by CuPc, where molecules form chain-like stacks and are each in the $S=1/2$ state due to the $Cu^{2+}$ ion. We have therefore chosen to fabricate and characterise CuPc thin films, and have extended our studies to CuPc crystals and MnPc to estimate the influence of structure and spin state on magnetic coupling.

Organic molecular beam deposition (OMBD) in an ultra-high vacuum chamber allows the creation of Pc thin films with a high degree of control and versatility. Figure 1.b shows how flexible films optimised for magnetic measurements can be grown on thermally stable polyimide substrates (kapton ®) by this method. The thickness of all films studied is typically 60nm, and the morphology of a CuPc film deposited on kapton at room temperature, Figure 1.d, shows small spherical crystallites. This is characteristic of planar phthalocyanines evaporated onto an amorphous substrate and is usually attributed to the so-called α-phase.[9] The polymorphic phase is confirmed by powder X-ray diffraction, and the corresponding unit cell is displayed in Figure 1.g. The molecules are arranged in a herringbone fashion, with their molecular planes parallel to each other within a column and forming an angle ϕ=65° with respect to the stacking direction. The films are highly textured with the (b,c) plane (projected onto the page) parallel to the substrate. Post-growth annealing of the α-CuPc film at 320°C leads to a phase transition to the β-polymorph, which is characterised by elongated crystallites (Figure 1.e), and a more compact structure with $\phi$ reduced to 45º, as seen from the unit cell represented in Figure 1.h. Since the intra-stack intermolecular separation is 3.4Å, the lateral shift between successive molecules in the β-phase is 3.4Å, more than twice the value of ~1.6Å for the α-phase. A third CuPc phase was created, by making use of the templating effect exerted by an underlying 20nm thick layer of PTCDA (3, 4, 9, 10-perylenetetracarboxylic dianhydride, Figure 1.c) which crystallises with its molecular planes parallel to the kapton substrate. The templating effect has been well-documented for glass/PTCDA/H$_2$Pc heterostructures.[10,11,12] and the templated phase consists of tilted α-phase columns, where the angle between the stacking axis and the surface has rotated from 0º to 65º, as illustrated in Figure 1.i. The CuPc molecules are therefore parallel to the substrate and shifted by 1.6Å with respect to each other. MnPc films were also formed by room temperature deposition on kapton substrates, and their structural and morphological features are comparable to the α-CuPc films, indicating that they also crystallise as the α-phase. Finally, single crystals of both CuPc and MnPc, which belong to the β-phase, were obtained by gradient sublimation.

Figure 2 exhibits the main results of magnetic measurements on Cu- and Mn-Pc films and single crystals (see details in the methods summary). Both the Curie-Weiss plots and magnetisation curves show clearly the switching of the magnetism between α and β polymorphs for Pc containing either transition metal. For the thin film α and templated phases, the Curie-Weiss plots have negative intercepts $\theta_p$ and the cross-over to full saturated magnetisation corresponds to large fields. This implies strong antiferromagnetic correlations which are only destroyed at elevated temperatures or fields. On the other hand, for both the thin film and bulk crystalline β-phases, we see that in the case of Cu, $\theta_p$ is close to zero, and the magnetisation curves are close to those for a Brillouin function for free S=1/2 ions with g=2. It can therefore be concluded that the magnetic couplings are mainly determined by the intermolecular shifts within columns, irrespective of physical appearance, crystal orientation, or inter-column interactions. These small shifts (<2Å) are able to modify the magnetic characteristics of the material, and the α → β phase transition corresponds to a conversion of the material from antiferro- to paramagnetic. The antiferromagnetic S=1/2 chains along the stacks are quantum antiferromagnets with strongly entangled spins, implying - more significantly - that a very straightforward annealing procedure switches the interactions of these interesting quantum objects.

For MnPc, related but somewhat different results are found – here, the tilting converts weaker antiferromagnetic into strongly ferromagnetic stacks, as can be seen immediately from the very steep magnetisation curve at T=2K, (red squares in Figure 2.d) and large and positive $\theta_p$=10K (Figure 2.c), in agreement with literature values.[4] In contrast, the alpha film (which was protected with a 100 nm thick layer of LiF in order to avoid oxidation of the manganese to a 3+ state) is very clearly antiferromagnetic, with $\theta_p$=-1.6K. The antiferromagnetic coupling is further confirmed by the rightwards deviation of the field-dependent magnetisation from a simulation of a paramagnetic spin =3/2 system using the Brillouin function (Figure 2.d, dashed line).

Table I shows the results of the Curie-Weiss analysis of the magnetisation for the different films, and clearly illustrates the polymorphic switching effects that are possible in the new nanocrystalline system of transition metal Pc films. For low-dimensional magnets, observed values for $\theta_p$ are proportional to, but actually somewhat larger than the exchange coupling; this follows because the Curie-Weiss form does not reproduce the well-known maximum eventually undergone by the susceptibility of the S=1/2 Heisenberg antiferromagnetic chain. To extract values for the coupling constant, $J_{exBF}$, appearing in the chain Hamiltonian H=$\Sigma_i$ $J_{exBF}$ $S_i S_{i+1}$, (where we neglect interchain interactions as well as the effects of potential dimerisations, both of which we could really only deduce beyond all doubt if we could perform inelastic neutron scattering, or low temperature STM, both of which are impossible propositions for our thin films on kapton) we performed a global fit to the Bonner-Fisher model[13] for all field-dependent and temperature-dependent magnetisations in the range of 0 to 7 Tesla and 2 to 30K. Figure 3.a shows the root mean squared deviation of the fits from the data as a function of $J_{exBF}$ for the different CuPc samples. The most likely values of $J_{exBF}$ can be simply read from the minima of the deviations, and are listed in Table I. The switching behaviour seen in the Curie-Weiss constants is reproduced, and the antiferromagnetic exchange constants $J_{exBF}$ are smaller than $\theta_p$, as expected.

We have shown that we can engineer magnetic couplings using polymorphism in transition metal Pc films. To understand the couplings in the different CuPc polymorphs, we have used *ab initio* calculations to determine the values of the exchange interactions from total energy differences. Absolute values were computed as a function of $\phi$ by comparing the energies of the singlet (estimated using the broken-symmetry approach developed by Noodelman[14]) and triplet states in a CuPc dimer, as calculated by Density Functional Theory (DFT) with the Gaussian code[15] and the B3LYP or UB3LYP exchange-correlation functionals,[16,17] which give good descriptions of exchange interactions in organic systems.[18] The blue triangles in Figure 3.b represent the results, and we can see that for $\phi$=45º and 65º, we obtain astonishingly good agreement between experiment and theory, i.e. the antiferromagnetic interaction is favoured in both cases, with J=-1.75K in the $\alpha$ phase, while the coupling is comparably negligible in the $\beta$ phase (J=-0.25K). One interesting finding from the calculations is the possibility of sign changes – i.e. switching between antiferro- and ferromagnetism - in the exchange interactions as a function of $\phi$, which occur between 30 and 40º.

While our results for the large polymorphism dependence of magnetic exchange are to our knowledge a spectacular success of DFT for organic matter, it is important to develop some intuition as to the origin of this effect, and in particular the possibility of a sign change in J as a function of $\phi$. The conventional mechanism for magnetic coupling between localised transition-metals involves superexchange [19] where an electron hops on and off the transition metal via the highest occupied or lowest unoccupied molecular orbitals (HOMO or LUMO). In CuPc, there are no MO's with the same symmetry as the Cu(II) ion ($b_{1g}$) near the Fermi edge, and superexchange must therefore be excluded. Instead, we propose that the magnetic coupling occurs via indirect exchange, [19] where the Pc system is transiently polarised by the direct Coulomb interaction with the Cu(II) spin, and this polarisation is then transferred to the neighbouring molecule by the hopping of a polarised electron-hole pair. This mechanism requires the electron and hole to occupy orbitals of the same symmetry, and the two $e_g$ levels just below and above the Fermi level satisfy this criterion. Perturbation theory to second order in the pair creation amplitude, and second order in the hopping matrix elements ($t_e$ and $t_h$ for electron and hole respectively, evaluated using the Kohn-Sham states obtained from DFT) leads to an exchange coupling which is directly proportional to $t_e t_h$. These products are scaled in order to match the experimental point at ϕ=65º, and are plotted in Figure 3.b as a function of ϕ. The interactions are weakly ferromagnetic at small ϕ, and negligible at ϕ=45º, consistent with the experimentally observed behaviour in β-CuPc. For increasing ϕ, the coupling then becomes antiferromagnetic, as observed in the negative J values obtained in α and templated CuPc. A qualitative understanding of the trends in $t_e t_h$ can be obtained from looking at the contributions to the overlap integrals for the orbitals involved, namely $e_g$ (HOMO) and $e_g$(LUMO), shown in Figure 3 in the x-polarisation. The mechanism implied in the perturbation theory is the molecular analogue of the Ruderman-Kittel-Kasuya-Yosida (RKKY) [20] interaction between localized moments in metals, and gives a qualitative explanation of the form of J, including the possibility of sign changes, but does not fully reproduce all features of the DFT calculations.

The theoretical treatment of the MnPc case is complicated by the interplay of various exchange mechanisms. Previous studies [8] have identified superexchange as the main path, but our calculations using perturbation theory suggest that superexchange should lead to structure-independent antiferromagnetic correlations, at odds with the strong ferromagnetism observed experimentally in the β-crystals. By using indirect exchange as for CuPc, we find that for ϕ=45º and 65º, the interactions are weakly ferromagnetic and strongly antiferromagnetic respectively, which provides a qualitative description of the system.

The polymorphic phthalocyanine films represent a new system where a molecular semiconductor is endowed with magnetic properties that can be switched with slight changes in crystal structure. Here we illustrate this process based on a simple annealing procedure, but we are investigating routes for reversible switching, using for example mechanical stress. Although the transition temperatures are still low, our theoretical estimates show that these can be improved with further crystal modifications and DFT will assist us to chose optimised materials and structures. Further enhancement in operation temperatures can be expected from using doping or more complex material systems (for example charge-transfer compounds) combined with our controlled growth methods. The high thermal stability of all phases below ~300ºC represents a departure from the usual bistable molecular systems where temperature is used to modify electronic level population and introduce lattice distortion.[21,22] Using local annealing, templating, and isomorphous substitution of different phthalocyanine derivatives, it will be possible to create new types of multilayer heterostructures which will provide a highly versatile ground for the development of new spintronic devices, without the requirements of epitaxy and compatibility which can restrict inorganic materials.

**Experimental**

Magnetic measurements were performed using a SQUID (Superconducting Quantum Interference Device)-based Magnetic Property Measurement System (MPMS) manufactured by Quantum Design. The temperature-dependent susceptibility presented is obtained from subtracting the magnetisation values at two fields (typically 1 and 2 Tesla), and dividing by their difference; the small diamagnetic contribution of the phthalocyanine rings themselves could be estimated from similar films and crystals of ZnPc and was subtracted from this data. For the single crystals, we have simply inserted them as powders into small gelatinous capsules and followed conventional protocol. The thin film measurements are much more challenging, and, exploiting the differential nature of the SQUID measurements, we have invented a new method to detect the low number (~$10^{16}$) of spins in the 60nm Pc films while eliminating any contribution from the substrate which is 25μm thick. This is achieved by depositing the film as a thin strip in the middle of a piece of kapton larger than the scanning distance (see Figure 1.b), and rolling the sample into the holder. The superposition of many rolls of film maximises the signal while the kapton, present in both the central measuring coils as well as the two background compensation coils of the magnetometer, makes no contribution at all.

**Acknowledgements**

SH thanks the Royal Society for a Dorothy Hodgkin research fellowship. Financial support from the Research Council UK and the Engineering and Physical Sciences Research Council (EPSRC) Basic Technology grant "Putting the Quantum into Information Technology" and a Royal Society Wolfson Research Merit Award is gratefully acknowledged.

**Figures and Table captions**

Figure 1 **Structure and morphology of phthalocyanine films**

(a) Structure of a metal phthalocyanine (MPc), (b) picture of a 2.5 cm$^2$ CuPc film deposited onto a 100 cm$^2$ kapton sheet, and (c) structure of PTCDA, used as a templating layer.  Atomic force micrographs of (d) a 60nm CuPc film deposited by OMBD at room temperature on kapton leading to α-phase crystallites, (e) a β-polymorph obtained after annealing for 2 hours at 320ºC and (f) a templated film deposited at room temperature onto a PTCDA first layer.  Schematics of the unit cells of (g) α-CuPc, (h) β-CuPc, and (i) templated CuPc, where $\phi$ is the angle between the stacking axis and the molecular planes.

Figure 2 **Magnetic properties of phthalocyanine films and crystals**

(a) Temperature dependent susceptibility on a range of CuPc films and crystals, with corresponding linear fits to the Curie-Weiss law (dashed line) and Bonner-Fisher fits (solid line). (b) Field dependent magnetisation for the CuPc films and crystals taken at 2K with resulting fits to the Bonner-Fisher model.  (c) Temperature dependent susceptibility on a α-MnPc film and MnPc crystals, with linear fits to the Curie-Weiss law, and (d) corresponding field dependent magnetisation measured at 2K for MnPc and simulation of the Brillouin function for a spin S = 3/2 system – strong crystal field effects and single ion anisotropy explain the slow increase of magnetisation above the value of 3 expected for a spin = 3/2 system,[23] and the differences in absolute magnetisations in the two polymorphs.[24]

Figure 3 **Calculated and measured values of exchange interactions in CuPc**

(Top half) The $e_{gx}$ molecular orbitals involved in the indirect exchange are shown for angles corresponding to both polymorphs and the maxima at $\phi$ = 45, 55, 65 and 90 º.  In this convention, a positive contribution to the hopping matrix elements between orbitals on the two molecules is shown in grey, while a negative contribution is shaded in black.  (a) Normalised sum of residuals between experimental data and Bonner-Fisher fits in the full space of temperatures and fields, for β-powder (continuous red), β-film (black dash-dot, divided by 5), α-film (blue dashed) and templated film (grey dotted). The normalised residuals, R, correspond to: $R = \sqrt{\dfrac{\sum (y_{ex} - y_{fit})^2}{N^2}}$ , where $y_{ex}$ and $y_{fit}$ are the experimental and fitted magnetisations respectively, and N is the number of data points. (b) Relative values of indirect exchange in CuPc dimers as a function of angle (red squares, normalised to fit experimental values), absolute values of exchange couplings, $J_{exDFT}$, obtained from DFT calculations (blue triangles), and experimental values of $J_{exBF}$ (grey circles).

Table 1 **Exchange interactions derived experimentally and theoretically**

Values of $\theta p$ extracted from fitting the experimental values of χ(T) to the Curie-Weiss law, $J_{exBF}$ obtained from fitting temperature and field dependent magnetisations to the Bonner Fisher model of isotropic antiferromagnetic chains and $J_{exDFT}$ estimated from *ab initio* calculations. Figure 2.a provides an estimate of the error on the values quoted.

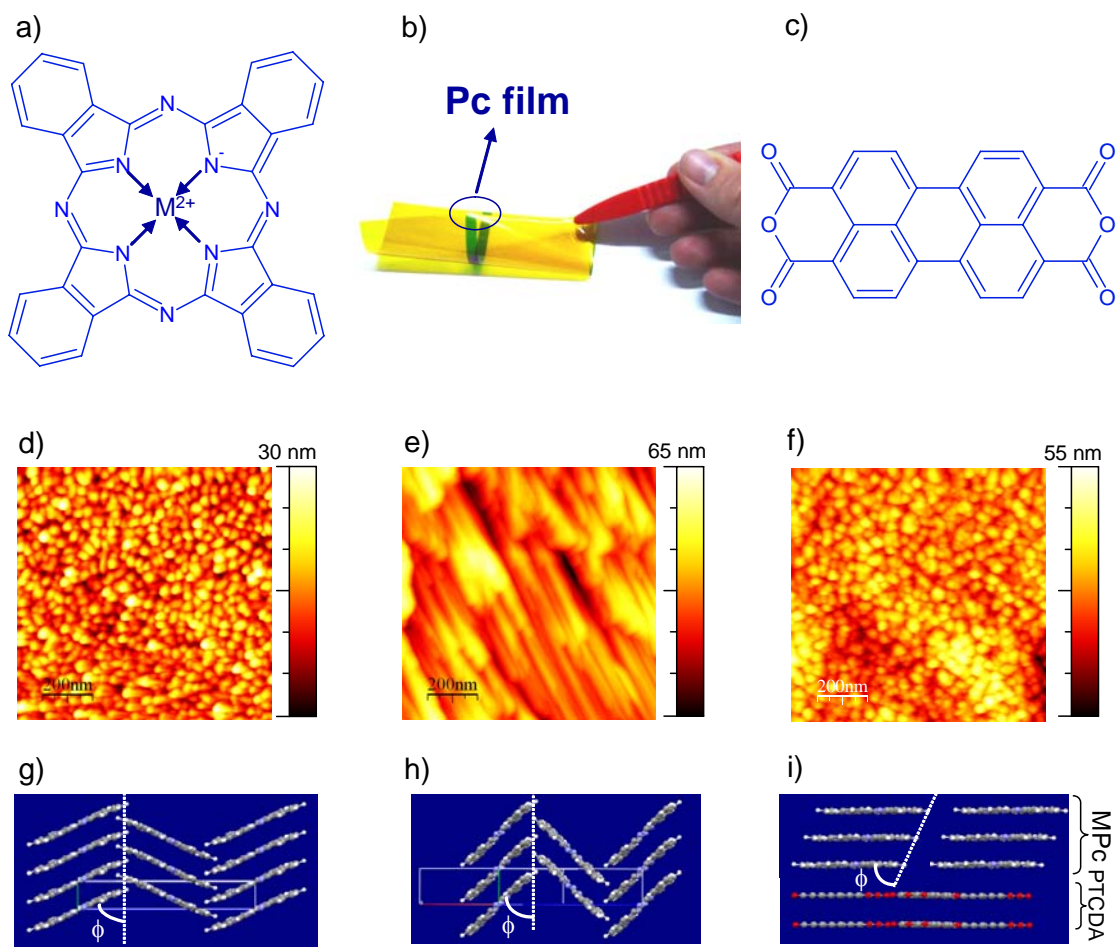

**Figure 1, Heutz *et al.***

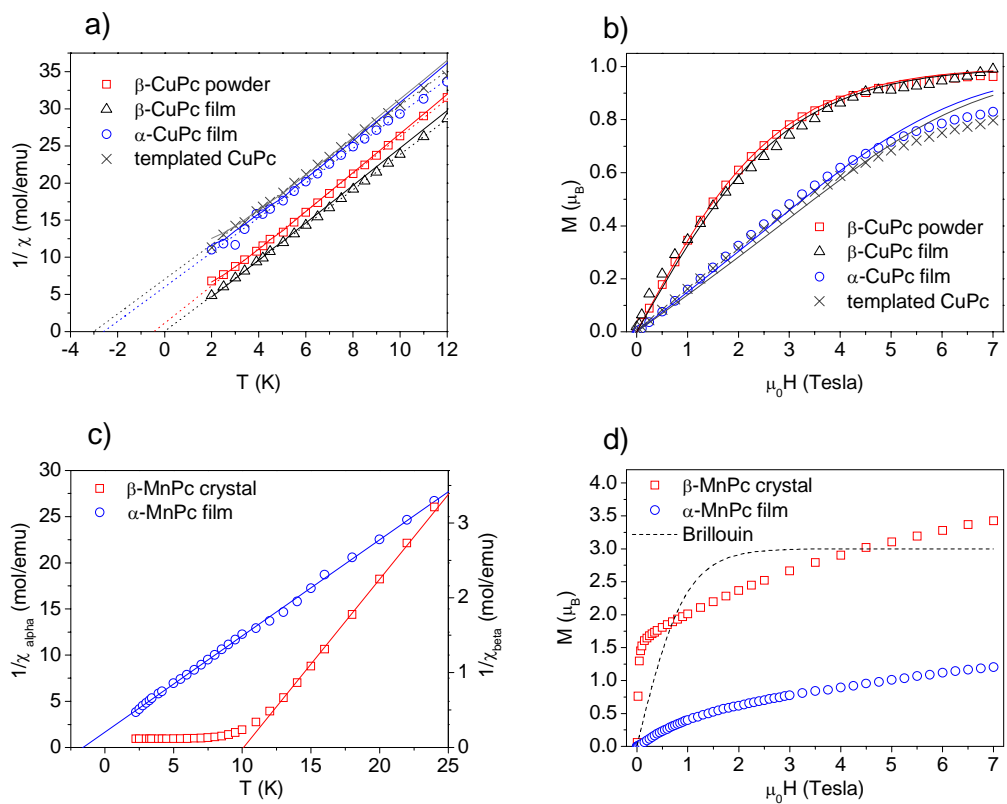

**Figure 2, Heutz *et al.***

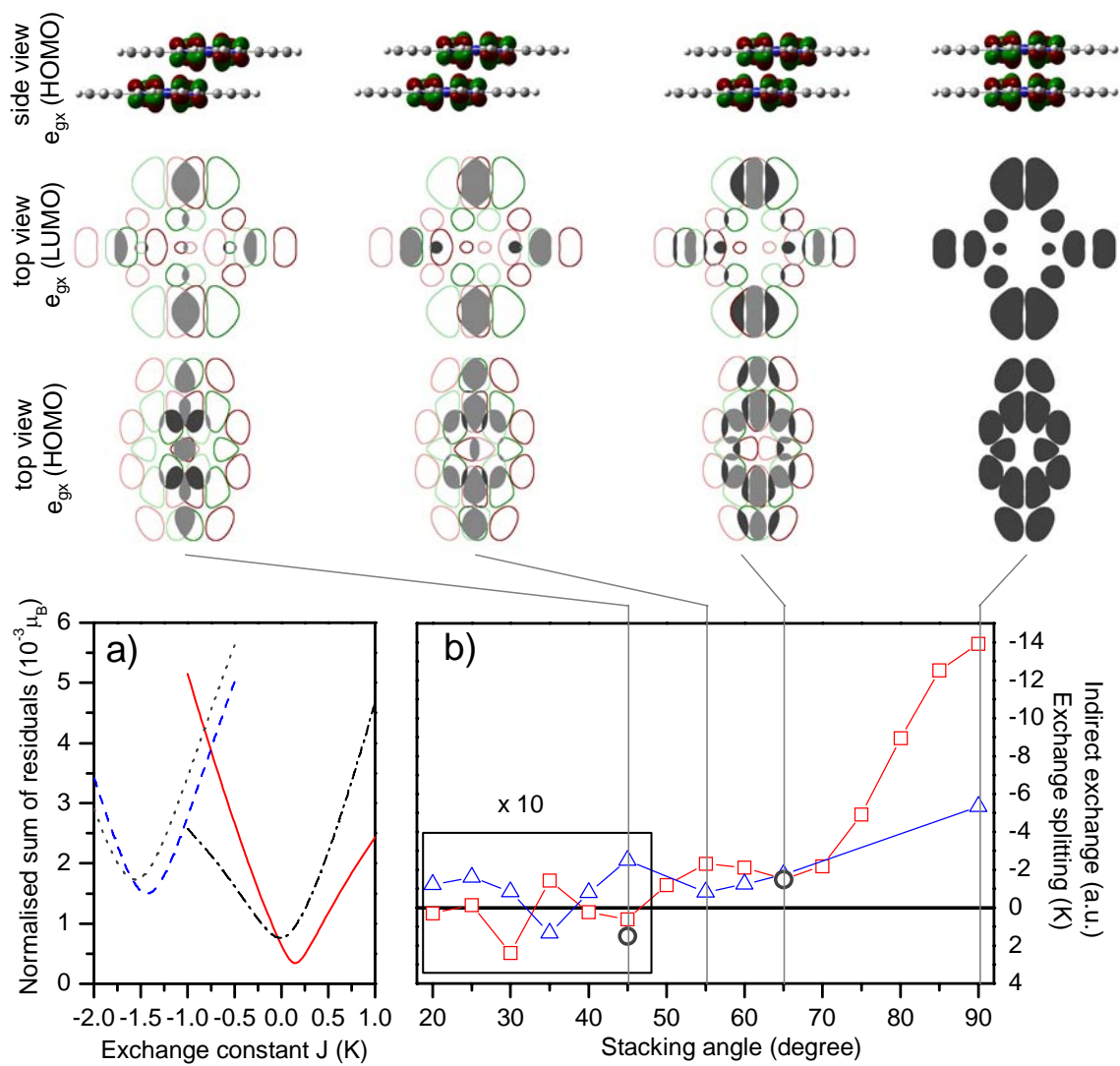

**Figure 3, Heutz *et al.***

|  | β-CuPc crystal | β-CuPc film | α-CuPc film | Templated CuPc film | β-MnPc crystal | α-MnPc film |
|---|---|---|---|---|---|---|
| $\theta_p$ / K | - 0.40 | 0.0 | - 2.6 | - 3.0 | 10 | -1.6 |
| $J_{exBF}$ / K | 0.15 | 0.0 | -1. 4 | -1.6 | / | / |
| $J_{exDFT}$ / K | - 0.25 | - 0.25 | -1.75 | -1.75 | / | / |

**Table 1, Heutz *et al.***